\documentclass{ws-p9-75x6-50}

\begin{document}

\title{Detecting a non-Gaussian stochastic background of gravitational radiation}

\author{Steve Drasco \& \'{E}anna \'{E}. Flanagan}
\address{Newman Laboratory of Nuclear Studies\\ Cornell University, Ithaca, New York 14853-5001, USA}  

\maketitle

\abstracts{
We derive a detection method for a stochastic background of gravitational waves produced
by events where the ratio of the average time between events to the average duration
of an event is large.  Such a signal would sound something like popcorn popping.  
Our derivation is based on the 
somewhat unrealistic assumption that the duration of an event is smaller than 
the detector time resolution. 
}

Consider a large collection of similar gravitational wave sources.  If
we cannot resolve the individual signals produced by these sources and 
know only their statistical properties, the signals
form a stochastic background. We group stochastic
backgrounds into  
the following two classes: those for which the individual sources
have (i) average redshift $z\gg1$  
and (ii) average redshift $z\sim 1$.  Typically in case (i) the number of 
individual sources is large enough that the duration of events is long
compared to the time between events, and then the central limit
theorem enforces Gaussianity of the background.  
However, for some sources in case (ii) the anticipated number of individual
sources might be small enough that this argument breaks down and a
non-Gaussian stochastic background might be expected.  
Examples of (ii) are supernovae\cite{Ferrari}, binary
inspirals, and young neutron stars.  Well developed data analysis
methods exist for detecting Gaussian backgrounds.  Here we derive a
data analysis method for a non-Gaussian background under idealized
assumptions; further work is needed to obtain realistic detection methods.
 
Assume a pair of co-located, aligned detectors with outputs
$h_i^k = n_i^k + s^k$ ($i=1,2$) of uncorrelated Gaussian noise $n_i^k$ and
possibly a stochastic background signal $s^k$.  Here $i$ labels the
detectors and $k$ with $1 \le k \le N$ labels the time samples of the
detector output. 
We assume the following probability distribution for the stochastic
background signal:
\begin{equation}\label{SignalProb}
p(\,{\bf s}\,|\xi,\alpha) = \prod_k \left\{
 \frac{\xi}{\sqrt{2\pi\alpha}} \exp \left[ -\frac{\left(s^k\right)^2}
{2\alpha}\right] + (1-\xi)\delta(s^k) \right\},
\end{equation}
where ${\bf s} = (s^1, \ldots, s^N)$.
Thus, the individual time samples are assumed to be statistically
independent and identically distributed; this is because we 
assume the duration of events is smaller than the detector sampling
time, so each event affects only one time sample\footnote{
Detector time resolutions can be on the order of milliseconds 
and even supernovae waveforms can be on the order of hundreds of
milliseconds, so this assumption is somewhat unrealistic.}.
In Eq.\ (\ref{SignalProb}), the parameter $\xi$ is the probability
that a randomly-chosen data sample will contain an event, and
$\alpha$ is the signal strength.  The Gaussian case is recovered in
the limit $\xi \to 1$.

We use Bayesian methods to motivate the choice of detection
statistic.  Let $P^{(1)}$ be the probability that a signal {\em is}
present in the data (posterior) and $P^{(0)}$ be the probability that
a signal {\em will be} present (prior).  The relation between these
probabilities is given in terms of the likelihood
ratio $\Lambda$ by 
\begin{equation}
\frac{P^{(1)}}{1 - P^{(1)}} = \Lambda 
\frac{P^{(0)}}{1-P^{(0)}}.
\end{equation}
A natural detection criterion is to threshold on $P^{(1)}$, which is
equivalent to thresholding on $\Lambda$.  Thus, we will say that the
signal has been detected if $\Lambda > \Lambda_*$ for some threshold
$\Lambda_*$\footnote{We restrict attention here to finding a useful
statistic and ignore the issue of how to choose the value of the
threshold, which may be done using either Bayesian or frequentist
methods.}.

To compute $\Lambda$ we need only detector data and the probability
density functions for the signal and for the noise in each detector.
We find
\begin{equation} \label{general}
\Lambda = \int d^Ns~
\frac{ 
                         p_{{\mathcal N}_1,{\mathcal N}_2} 
                         ({\bf h}_1 - {\bf s}
                          ,{\bf h}_2- {\bf s} ) 
     }
     {
                        p_{{\mathcal N}_1,{\mathcal N}_2}
                        ({\bf h}_1 ,{\bf h}_2) 
     } 
 \tilde p^{(0)}_{{\mathcal S}}({\bf s}),
\end{equation}
where ${\bf h}_i$ is the vector $(h_i^1, \ldots,
h_i^N)$, $p_{{\mathcal N}_1,{\mathcal N}_2}
({\bf n}_1,{\bf n}_2)$ is the joint probability density of the
detector noises ${\bf n}_1$ and ${\bf n}_2$, 
and  
$\tilde p^{(0)}_{{\mathcal S}}({\bf s})$ is the probability density
that the signal ${\bf s}$ is present in the data, {\em given that some
signal is in the data}.  We take 
$\tilde p^{(0)}_{{\mathcal S}}({\bf s}) = \int d\xi~ \int d\alpha~
p({\bf s}|\xi,\alpha) \tilde p^{(0)}(\xi,\alpha)$, where 
$\tilde p^{(0)}(\xi,\alpha)$ is the prior distribution for the signal
parameters $\xi$ and $\alpha$.  Combining these equations and assuming
uncorrelated white Gaussian noise of unit variance in each detector\footnote{We
expect to be able to relax this assumption in the near future.} we obtain
\begin{equation}
\Lambda = \int d\xi~ \int d\alpha~
\Lambda(\xi,\alpha)
\tilde p^{(0)}(\xi,\alpha)
\end{equation}
where 
\begin{equation}\label{result}
\ln \Lambda \left(\xi,\alpha \right) = \sum_j  \ln \left( 1 + \xi v^j \right) \nonumber \\
\end{equation}
and
\begin{equation}
v^j   \equiv \left(1 + 2\alpha \right)^{-1/2} \exp 
             \left[ \left( h_1^j + h_2^j\right)^2
             \frac{\alpha}{4\alpha + 2} \right] - 1. 
\end{equation}

The detection statistic we suggest is the maximum likelihood statistic
\begin{equation}
\Lambda_{\rm max} = \max_{\alpha,\xi} \ \Lambda(\xi,\alpha),
\end{equation}
which we expect to be large when $\Lambda$ itself is large.  
The maximization can be carried out numerically\footnote{While such a
search is computationally intensive, we have had some success with
numerical simulations using small (length $\sim 10^6$) data
segments.}.  If a signal is detected, then the values $\xi_{\rm max}$
and $\alpha_{\rm max}$ which achieve the maximum give estimators of
the true values of $\alpha$ and $\xi$.  One can also form a posterior
distribution for $\alpha$ and $\xi$ from the function
$\Lambda(\alpha,\xi)$ and thereby attempt to distinguish between Gaussian and
non-Gaussian backgrounds.

\end{document}